\documentclass [a4paper,fleqn,12pt]{article}
\usepackage{graphicx}
\usepackage {subfigure,epsfig}
\usepackage {amssymb,amsmath,latexsym,enumerate,graphicx}
\usepackage{cite}
\usepackage{amsthm}
\numberwithin{equation}{section} \numberwithin{table}{section}
\mathindent=0pt \theoremstyle{plain} \newtheorem{theorem}{Theorem}
\newtheorem{lemma}{Lemma}

\numberwithin{theorem}{section} \numberwithin{lemma}{section}

\begin{document}

\title{\textbf{The Yablonskii - Vorob'ev polynomials for the second Painlev\'{e} hierarchy}}

\author{Maria V. Demina, Nikolai A. Kudryashov}

\date{Department of Applied Mathematics\\
Moscow Engineering and Physics Institute\\ (State University)\\
31 Kashirskoe Shosse, 115409, Moscow, \\ Russian Federation}
\maketitle

\begin{abstract}
Special polynomials associated with rational solutions of the second
Painlev\'{e} equation and other equations of its hierarchy are
studied. A new method, which allows one to construct each family of
polynomials is presented. The structure of the polynomials is
established. Formulaes for their coefficients are found. The degree
of every polynomial is obtained. The main achievement of the method
lies in the fact that it enables one to construct the family of
polynomials corresponding to any member of the second Painlev\'{e}
hierarchy. Our approach can be applied for deriving the polynomials
related to rational or algebraic solutions of other nonlinear
differential equations.
\end{abstract}

\emph{Keywords:} the Yablonskii - Vorob'ev polynomials, the second
Painlev\'{e} equation,  the second Painlev\'{e} hierarchy, power geometry\\

PACS: 02.30.Hq - Ordinary differential equations

\section{Introduction}

It is well known that the second Painlev\'{e} equation $(P_2)$
\begin{equation}
\label{1.1}w_{zz}=2w^3+zw+\alpha
\end{equation}
and all members of its hierarchy have rational solutions only at
integer values of the parameter $\alpha$ $(\alpha=n\in \textbf{Z})$.
These solutions can be written in terms of special polynomials
$Q_n^{(N)}(z)$ $(N\geq1)$
\begin{equation}
\begin{gathered}
\label{1.2}\hfill w^{(N)}(z;n)=\frac{d}{dz}\left\{\ln
\left[\frac{Q_{n-1}^{(N)}(z)}{Q_n^{(N)}(z)}\right]\right\},\quad\,
n\geq1,\quad\,\,w^{(N)}(z;-n)=-w^{(N)}(z;n),
\end{gathered}
\end{equation}
where $w^{(N)}(z;n)$ is a solution of the Nth equation in the
hierarchy (N=1 is the case of $P_2$). The polynomials $Q_n^{(1)}(z)$
were suggested by A. I. Yablonskii and A. P. Vorob'ev and are called
the Yablonskii -- Vorob'ev polynomials \cite{Yablonskii01,
Vorob'ev01}. While their analogues were introduced by P. A. Clarkson
and E. L. Mansfield \cite{Clarkson01}. All these polynomials can be
regarded as nonlinear analogues of classical special polynomials
\cite{Clarkson02, Clarkson03, Clarkson04, Okamoto01, Umemura01,
Nuomi01, Kajiwara01, Fucutani01, Taneda01}.

The polynomials $Q_n^{(1)}(z)$ satisfy the differential --
difference equation
\begin{equation}
\label{1.3}Q_{n+1}^{(1)}Q_{n-1}^{(1)}=z(Q_n^{(1)})^2
-4[Q_n^{(1)}(Q_n^{(1)})^{''}-(Q_n^{(1)})^{'}(Q_n^{(1)})^{'}],
\end{equation}
where $Q_0^{(1)}(z)=1$, $Q_1^{(1)}(z)=z$. It is not clear from the
first sight that this relation defines exactly polynomials however
it is so. Moreover $Q_n^{(1)}(z)$ are monic polynomials with integer
coefficients. They possess a certain number of interesting
properties. In particular, for every integer positive $n$ each
polynomial $Q_n^{(1)}(z)$ has simple roots only and besides that,
two successive polynomials $Q_n^{(1)}(z)$ and $Q_{n+1}^{(1)}(z)$ do
not have a common root. The Yablonskii -- Vorob'ev polynomials
$Q_n^{(1)}(z)$ arise in various physical models. For example,
partial solutions of the Korteveg -- de Vries equation, the modified
Korteveg -- de Vries equation, the nonlinear Schr\"{o}dinger
equation, the Kadomtsev -- Patviashvili equation can be expressed
via the polynomials $Q_n^{(1)}(z)$ \cite{Clarkson02}.

The polynomials $Q_n^{(2)}(z)$ associated with the second equation
of the $P_2$ hierarchy
\begin{equation}
\label{1.4a}w_{zzzz}-10\,w^2\,w_{zz}-10\,w\,w_z^2+6\,w^5-z\,w-\alpha=0
\end{equation}
in their turn satisfy the differential -- difference equation
\begin{equation}
\begin{gathered}
\label{1.5}Q_{n+1}^{(2)}Q_{n-1}^{(2)}=z(Q_n^{(2)})^2-4\left[Q_n^{(2)}
(Q_n^{(2)})^{''''}-4(Q_n^{(2)})^{'''}(Q_n^{(2)})^{'}\right.\\
\left.+3(Q_n^{(2)})^{''}(Q_n^{(2)})^{''}\right],
\end{gathered}
\end{equation}
where again $Q_0^{(2)}(z)=1$, $Q_1^{(2)}(z)=z$. As far as the
polynomials $Q_n^{(N)}(z)$ $(N\geq3)$ are concerned, the
differential -- difference equation analogues to \eqref{1.3},
\eqref{1.5} yet is not obtained. In this connection an important
problem is to define the polynomials in another way, i.e. without
using the differential -- difference equation.

In this paper we present a new method, which solves this problem. In
particular, the method allows one to find the degree of each
polynomial $Q_n^{(N)}(z)$, to determine its structure and to derive
formulas for its coefficients. Our approach can be also applied for
constructing other polynomials related to rational or algebraic
solutions of nonlinear differential equations. We would like to
mention that not long ago it was made an attempt to obtain formulas
for coefficients of the Yablonskii - Vorob'ev polynomials
$Q_n^{(1)}(z)$ \cite{Kaneko01}. As a result the coefficient of the
lowest degree term was found.

The outline of this paper is as follows. In section 2 the algorithm
of our method is presented and main theorems are proved.
Correlations between the roots of the Yablonskii - Vorob'ev
polynomials and properties of their coefficients are discussed in
sections 3 and 4, accordingly. In other words sections 2 - 4 are
devoted to the general case. While several examples are given in
sections 5 - 7. More exactly the polynomials associated with the
first, the second, and the third members of the hierarchy are
studied in sections 5, 6, 7, respectively.

\section{Method applied}

The $P_2$ hierarchy can be obtained through the scaling reduction
from the hierarchy of the modified Korteveg -- de Vries equation
\cite{Kudryashov01, Kudryashov02, Kudryashov02a, Kudryashov02b,
Hone01}, and is the following

\begin{equation}
\label{2.1}P_2^{(N)}[w,\alpha]=(\frac{d}{dz}+2w)L_N[w^{'}-w^2]-zw-\alpha=0,\quad
N\geq1,
\end{equation}
where $L_N[w^{'}-w^2]\equiv L_N[u]$ satisfies the Lenard recursion
relation

\begin{equation}
\label{2.2}d_zL_{N+1}[u]=(d^3_z+4ud_z+2u_z)L_N[u],\quad
L_0[u]=\frac12.
\end{equation}

Every equation of the hierarchy \eqref{2.1} has a unique rational
solution if and only if $\alpha$ is an integer. All these solutions
are expressible via the logarithmic derivative of the polynomials
$Q_n^{(N)}(z)$ $(N\geq1)$, which will be the objects of our study.
Analyzing the expression \eqref{1.2} we understand that the
polynomials $Q_n^{(N)}(z)$ can be defined as monic polynomials. By
$p_{n,N}$ denote the degree of $Q_n^{(N)}(z)$. Then we can present
each polynomial in the form

\begin{equation}
\label{2.3}Q^{(N)}_n(z)=\sum_{k=0}^{p_{n,N}}A^{(N)}_{n,k}z^{p_{n,N}\,\,-\,\,k},\qquad
A^{(N)}_{n,0}=1
\end{equation}

Let us show that it is possible to derive the polynomials
$Q_n^{(N)}(z)$ without leaning on the differential -- difference
recurrence formula or determinantal representation of the rational
solutions. For this aim we will use power expansions at infinity for
solutions of the equations \eqref{2.1}.
\begin{theorem}
\label{T:1} Every equation $P_2^{(N)}[w,\alpha_N]$ has a family of
solutions with the expansion at infinity

\begin{equation}
\label{2.4}w^{(N)}(z;\alpha)=-\frac{\alpha}{z}+\sum_{l=1}^{\infty}
c_{\alpha,-(2N+1)l-1}z^{-(2N+1)l-1},
\quad z\rightarrow \infty,\quad N\geq1.
\end{equation}

\end{theorem}

\begin{proof}
While proving this theorem we will use the algorithms of power
geometry. For more information see \cite{Bruno01, Bruno02, Bruno03,
Kudryashov03, Demina01, Kudryashov04}. First of all it is important
to mention that equations \eqref{2.1} and the operators
$L_N[w^{'}-w^2]$ can be thought of as differential sums with $z$
being an independent variable and $w$ being a dependent one. Let us
show that the support $S(L_N)$ of $L_N[w^{'}-w^2]$ satisfies the
correlation

\begin{equation}
\begin{gathered}
\label{2.5}\{(-2N+1,1);\,(0,2N)\}\subset S(L_N)\subseteq
\{(-2N+m,m); 1\leq m\leq2N\}.
\end{gathered}
\end{equation}

\begin{figure}[h]
 \centerline{
 \subfigure[]{\epsfig{file=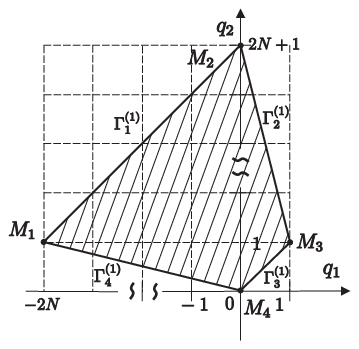,width=70mm}\label{fig1:z_post_1}}
 \subfigure[]{\epsfig{file=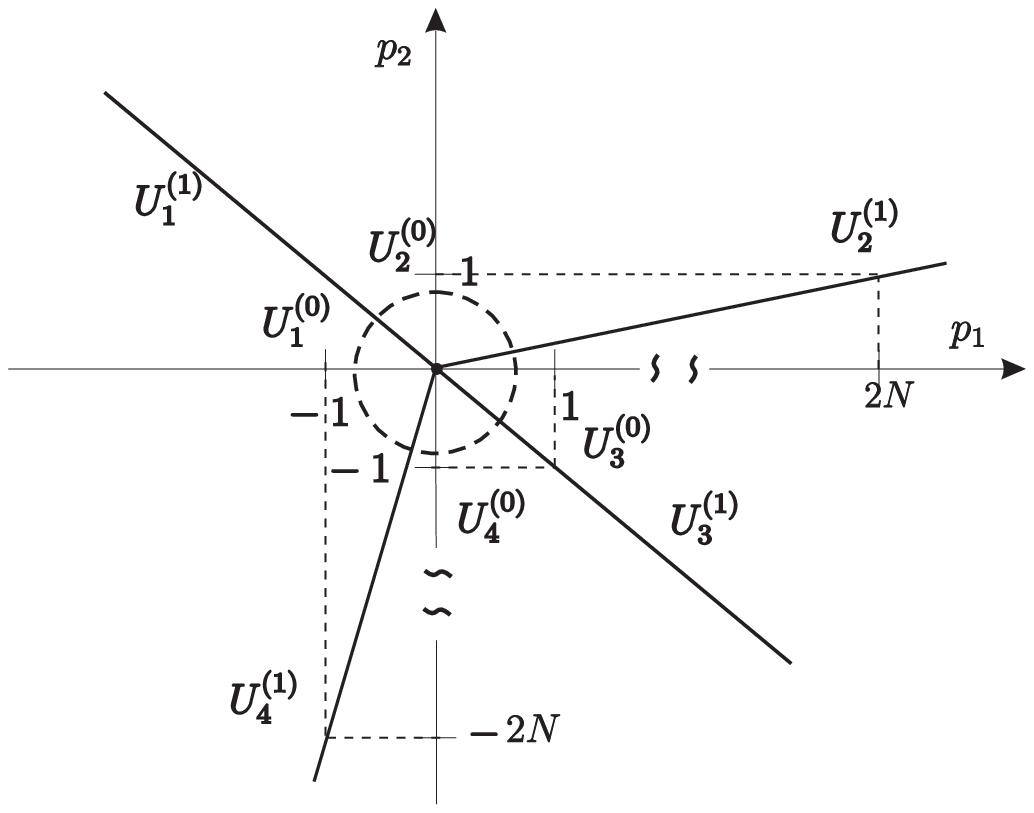,width=90mm}\label{fig2:z_post_2}}}
 \caption{\textbf{a}: Polygon corresponding to the Nth equation in the $P_2$
 hierarchy; \textbf{b}: The normal cones for the vertexes
 $\Gamma_j^{(0)}\equiv M_j$ and for the edges $\Gamma_j^{(1)}$\,(j=1,2,3,4).}
 \label{fig1:z_post}
\end{figure}

This fact can be proved by induction. For $N=1$ it is obvious.
Suppose it is true for $N=M$. Using the recurrent expression
\eqref{2.2} we get

\begin{equation*}
\begin{gathered}
\label{2.6}\{(-2(M+1)+1,1)\}\subset S(L_{M+1})\subseteq
\{(-2M+m,m)+(-2,0); 1\leq m\leq2M\}\cup\\
\{(-2M+m,m)+(-1,1); 1\leq m\leq2M\}\cup  \{(-2M+m,m)+(0,2); 1\leq
m\leq2M\}.
\end{gathered}
\end{equation*}

Thus we immediately obtain

\begin{equation}
\begin{gathered}
\label{2.6}\{(-2(M+1)+1,1)\}\subset S(L_{M+1})\subseteq
\{(-2(M+1)+m,m); \\
1\leq m\leq2M\}\cup \{(-2(M+1)+m,m); 2\leq m\leq2M+1\}\cup\\
\{(-2(M+1)+m,m); 3\leq m\leq2(M+1)\}.
\end{gathered}
\end{equation}

Deriving this expression we did not pay attention to coefficients at
the monomials of $L_{M+1}[w^{'}-w^2]$. The monomial $w^{2M}$ of
$L_M[w^{'}-w^2]$ becoming the monomial $w^{2(M+1)}$ of
$L_{M+1}[w^{'}-w^2]$ may disappear. However, this does not happen as
the coefficient at $w^{2M}$ in $L_M[w^{'}-w^2]$ is equal to
$(-1)^M2^{M-1}(2M-1)!!/M!\neq 0$, $M\geq1$. Hence we get

\begin{equation}
\begin{gathered}
\label{2.7}\{(-2(M+1)+1,1);\,(0,2(M+1))\}\subset
S(L_{M+1})\subseteq\\
 \{(-2(M+1)+m,m); 1\leq m\leq2(M+1)\}.
\end{gathered}
\end{equation}

Using the expression \eqref{2.1} it can be easily proved that
\begin{equation}
\begin{gathered}
\label{2.8}\{(-2N,1);\,(0,2N+1);(1,1);(0,0)\}\subset
S(P_2^{(N)})\subseteq \\
\{(-2N+m,m+1);(1,1);(0,0); 0\leq m\leq2N\},
\end{gathered}
\end{equation}
where $S(P_2^{(N)})$ stands for the support of the Nth equation in
the hierarchy. From the previous expression we see that the polygon
corresponding to the equation $P_2^{(N)}$ at $\alpha\neq0$ is a
trapezium with the vertexes $M_1=(-2N,1)$, $M_2=(0,2N+1)$,
$M_3=(1,1)$, $M_4=(0,0)$ and with the edges
$\Gamma_1^{(1)}=[M_1,M_2]$, $\Gamma_2^{(1)}=[M_2,M_3]$,
$\Gamma_3^{(1)}=[M_3,M_4]$, $\Gamma_4^{(1)}=[M_1,M_4]$ (see Fig. 1).
The support of $P_2^{(N)}$ lies in the lattice $\textbf{Z}$ with the
basis $B_1=(-2N,1)$, $B_2=(1,1)$ (again $\alpha\neq0$). Let us find
the power expansion corresponding to the edge $\Gamma_3^{(1)}$. This
edge is characterized by the reduced equation

\begin{equation}
\label{2.8a}\hat{f}_3^{(1)}\stackrel{def}{=} -zw-\alpha=0,\quad (
f(z,w)\equiv P_2^{(N)}(z,w))
\end{equation}

and the normal cone $U^{(1)}_3=\{\lambda(1,-1)$, $\lambda>0\}$.
Hence the power asymptotics related to the edge $\Gamma_3^{(1)}$ is
the following

\begin{equation}
\label{2.9}w^{(N)}(z;\alpha)\sim-\frac{\alpha}{z},\quad z\rightarrow
\infty,\quad N\geq1
\end{equation}

The reduced equation \eqref{2.8a} is algebraic, therefore its
solution does not have critical numbers. A shifted support of
\eqref{2.9} is the vector $B=(-1,-1)=-B_2$. Consequently it belongs
to the lattice $\textbf{Z}$ generated by the vectors $B_1$, $B_2$.
This lattice consists of the points $M=\{(q_1,q_1)$, $q_1=-2Nl+m$,
$q_2=l+m\}$, where $m$ and $l$ are whole numbers. The lattice
$\textbf{Z}$ intersects with the line $q_{2}=-1$ by the points
$q_1=-(2N+1)m-1$. The cone of the problem is $\mathcal{K}=\{k<-1\}$.
Thus the exponents of the power expansion with the asymptotic
behavior \eqref{2.8} belong to the set $\textbf{K}=\{k=-(2N+1)m-1$,
$m\in \mathbb{N}\}$. This completes the proof.
\end{proof}
The expansion \eqref{2.4} at $N=1$ and $N=2$ was found in
\cite{Bruno01} and \cite{Demina01}, accordingly. All the
coefficients $c_{\alpha,-(2N+1)l-1}$ $(l\geq1)$ in \eqref{2.4} can
be sequently calculated. For convenience of use let us present the
series \eqref{2.4} in the form

\begin{equation}
\label{2.11}w^{(N)}(z;\alpha)=\sum_{m=0}^{\infty}c_{\alpha,-(m-1)}z^{-m-1},
\end{equation}
where $c_{\alpha,-(m-1)}=0$ unless $m$ is divisible by $(2N+1)$.

The first two solutions of the equation $P_2^{(N)}$ are the
following $w^{(N)}_0=0$, $w^{(N)}_1=-1/z$. This fact can be verified
by direct substitution into \eqref{2.1}. Therefore it can be set
$Q^{(N)}_0=1$, $Q^{(N)}_1=z$. Suppose $a^{(N)}_{n,k}$ $(1\leq k\leq
p_{n,N})$ are the roots of the polynomial $Q^{(N)}_n(z)$, then by
$s^{(N)}_{n,k}$ we denote the symmetric functions of the roots

\begin{equation}
\label{2.12}s^{(N)}_{n,m}\stackrel{def}{=}\sum_{k=1}^{p_{n,N}}(a^{(N)}_{n,k})^m,\quad
m\geq 1.
\end{equation}

Later in this section when it does not cause any contradiction the
index $N$ will be omitted. Our next step is to express $s_{n,m}$
through coefficients of the series \eqref{2.11}.

\begin{theorem}
\label{L:1} Let $c_{i,-m-1}$ be the coefficient in expansion
\eqref{2.11} at integer $\alpha=i\in \textbf{N}$. Then for each
$n\geq2$ the following relations hold

\begin{equation}
\label{2.13}s_{n,m}=-\sum_{i=2}^{n}c_{i,-(m+1)},\quad m\geq1,
\end{equation}

\begin{equation}
\label{2.13a}p_n=-\sum_{i=1}^{n}c_{i,-1}.
\end{equation}

\end{theorem}

\begin{proof}
As far as $Q_n(z)$ is a monic polynomial, then it can be written in
the form

\begin{equation}
\label{2.14}Q_n(z)=\prod_{k=1}^{p_{n}}(z-a_{n,k}).
\end{equation}

Note that possibly $a_{n,k}=a_{n,l}$, $k\neq l$. This equality
implies that

\begin{equation}
\label{2.15}\frac{Q_n^{'}(z)}{Q_n(z)}=\sum_{k=1}^{p_{n}}\frac1{z-a_{n,k}}.
\end{equation}

Substituting \eqref{2.15} into the expression \eqref{1.2} yields

\begin{equation}
\label{2.16}w(z;n)=
\sum_{k=1}^{p_{n-1}}\frac1{z-a_{{n-1},k}}-\sum_{k=1}^{p_{n}}\frac1{z-a_{n,k}}.
\end{equation}

Expanding this function in a neighborhood of infinity we get

\begin{equation}
\begin{gathered}
\label{2.17}w(z;n)= \frac{b_{n-1}}{z}-\frac{b_n}{z}+
\sum_{m=0}^{\infty}
\left[\sum_{k=1+b_{n-1}}^{p_{n-1}}(a_{n-1,k})^m-\sum_{k=1+b_n
}^{p_{n}}(a_{n,k})^m\right]\\
\times z^{-(m+1)},\,\, |z|>\max\{\tilde{a}_{n-1},\tilde{a}_{n}
\},\,\, \tilde{a}_{n}=\max\limits_{1\leq k \leq
p_{n}}\{|a_{n,k}|\},\,\,b_n=\sum_{k=1}^{p_{n}}\delta_{0,a_{n,k}},
\end{gathered}
\end{equation}
where $\delta_{0,a_{n,k}}$ is the Kronecker delta. The first or the
second term in \eqref{2.17} are present only if the polynomials
$Q_{n-1}(z)$, $Q_n(z)$ have zero roots, accordingly. In our
designations the previous expression can be rewritten as

\begin{equation}
\begin{gathered}
\label{2.18}w(z;n)=-\frac{p_{n}-p_{n-1}}{z}+\sum_{m=1}^{\infty}\left[s_{n-1,m}-
s_{n,m}\right]z^{-(m+1)},\\
|z|>\max\{\tilde{a}_{n-1},\tilde{a}_{n}\}.
\end{gathered}
\end{equation}

The absence of a zero term in sum is essential only at $m=0$.
Comparing expansions \eqref{2.18} and \eqref{2.11} we obtain the
equalities

\begin{equation}
\begin{gathered}
\label{2.19}
p_{n}-p_{n-1}=-c_{n,-1},\\
s_{n,m}-s_{n-1,m}=-c_{n,-(m+1)},\quad m\geq1.
\end{gathered}
\end{equation}

Decreasing the first index by one in \eqref{2.19} and adding the
result to the original one yields

\begin{equation}
\begin{gathered}
\label{2.20}p_{n}-p_{n-2}=-(c_{n,-1}+c_{n-1,-1}),\\
s_{n,m}-s_{n-2,m}=-(c_{n,-(m+1)}+c_{n-1,-(m+1)}).
\end{gathered}
\end{equation}

Note that $c_{1,-(m+1)}=0,\,m\geq1$ and $a_{1,1}=0$. Then proceeding
in such a way we get the required relations \eqref{2.13} and
\eqref{2.13a}.

\end{proof}

\textit{Remark 1.} It was proved that $P_2^{(N)}$ has a unique
rational solution whenever $\alpha$ is an integer. All these
solutions possess convergent series at infinity. For $n\neq0$ every
rational solution $w^{(N)}(z;n)$ has the asymptotic behavior

\begin{equation*}
\label{2.21}w^{(N)}(z;n)\backsim -\frac{n}{z},\quad z\rightarrow
\infty,
\end{equation*}
i.e. the point $z=\infty$ is a simple root. This fact can be easily
seen from the B\"{a}cklund transformation for $P_2^{(N)}$. Thus the
formal series \eqref{2.4} at $\alpha=n$ coincides with the expansion
\eqref{2.18} and is also convergent.

\textit{Remark 2.} Since $c^{(N)}_{i,-1}=-i$ $\forall N\geq1$, we
get that the degree of each polynomial $Q_n^{(N)}(z)$ is

\begin{equation}
\label{2.21}p_{n,N}=\sum_{i=1}^{n}i=\frac{n(n+1)}{2}.
\end{equation}

Theorem \eqref{L:1} enables us to prove the following theorem.
\begin{theorem}
\label{T:1.1.} All the coefficients $A_{n,m}$ of the polynomial
$Q_n(z)$ can be obtained with a help of $n(n+1)/2+1$ first
coefficients of the expansion \eqref{2.11} for the solutions of
$P_2^{(N)}$.

\end{theorem}

\begin{proof} For every polynomial there exists a connection between
its coefficients and the symmetric functions of its roots $s_{n,m}$.
This connection is the following

\begin{equation}
\label{2.22}mA_{n,m}+s_{n,1}A_{n,m-1}+\ldots +s_{n,m}A_{n,0}=0,\quad
1\leq m\leq p_n.
\end{equation}

Taking into account that in our case $A_{n,0}=1$ and $p_n=n(n+1)/2$
we get

\begin{equation}
\label{2.23}A_{n,m}=-\frac{s_{n,m}+s_{n,m-1}A_{n,1}+\ldots
+s_{n,1}A_{n,m-1}}{m},\quad 1\leq m\leq n(n+1)/2.
\end{equation}

The function $s_{n,m}$ can be derived using the expression
\eqref{2.13}. Hence recalling the fact that \eqref{2.11} is exactly
\eqref{2.4} we obtain

\begin{equation}
\begin{gathered}
\label{2.24}s_{n,m}=0,\quad m\in \textbf{N}\,/\,\{(2N+1)l,\quad l\in
\textbf{N}\},\\
s_{n,(2N+1)\,l}=-\sum_{i=2}^{n}c_{i,-(2N+1)\,l-1},\quad l\in
\textbf{N}.
\end{gathered}
\end{equation}

Substituting this into \eqref{2.23} yields

\begin{equation}
\begin{gathered}
\label{2.25}A_{n,m}=0,\quad m\in\{1,2,\ldots
,n(n+1)/2\}\,/\,\{(2N+1)l,\quad l\in \textbf{N}\};\\
A_{n,(2N+1)\,l}=-\frac{1}{(2N+1)l}\{s_{n,(2N+1)\,l}+s_{n,(2N+1)\,l-(2N+1)}A_{n,(2N+1)}+\\
\ldots +s_{n,(2N+1)}A_{n,(2N+1)\,l-(2N+1)}\},\quad l\in
\textbf{N},\,(2N+1)l\leq n(n+1)/2.
\end{gathered}
\end{equation}

Thus we see that the coefficients $A_{n,k}$ of the polynomial
$Q_n(z)$ are uniquely defined by coefficients $c_{n,-{(2N+1)l}-1}$
of the expansion \eqref{2.4}. This completes the proof.
\end{proof}
\textit{Remark 3.} Expression \eqref{2.25} defines the structure of
the polynomial $Q_n(z)$. Namely if $n(n+1)/2$ is divisible by
$(2N+1)$, i.e. $n\equiv 0\,mod\, (2N+1)$ or $n\equiv2N\,
mod\,(2N+1)$, then $Q_n(z)$  is a polynomial in $z^{2N+1}$.
Otherwise (if $n(n+1)/2$ is not divisible by $(2N+1)$) $Q_n(z)/z^r$
is a polynomial in $z^{2N+1}$, where $r=n(n+1)/2\,mod\,(2N+1)$. In
other words

\begin{equation}
\begin{gathered}
\label{2.26}q_n^{(N)}(z)\stackrel{def}{=}\frac{Q_n^{(N)}
(\xi^{1/(2N+1)})}{\xi^{r/(2N+1)}},\quad
\xi=z^{2N+1}
\end{gathered}
\end{equation}
is a polynomial of degree $[n(n+1)/(4N+2)]$ with $[x]$ denoting the
integer part of $x$.

\section{Symmetric functions of the roots}

In this section we are discussing properties of the symmetric
functions. It is important to note that the functions
$s_{n,m}^{(N)}$ can be regarded as relations between the roots
$a_{n,k}^{(N)}$ of the polynomials $Q^{(N)}_n(z)$. Recently the
location of the roots in the complex plane was investigated. It was
shown that the structure of the roots is very regular. In order to
establish our main results we need a lemma.
\begin{lemma}
\label{L:3.1} The coefficient $c_{\alpha,-(2N+1)l-1}$ in the
expansion \eqref{2.4} is a polynomial in $\alpha$ of degree $2Nl+1$.
\end{lemma}
\begin{proof}
The proof is by induction on $l$. For $l=0$ there is nothing to
prove as $c_{\alpha,-1}=-\alpha$. Other coefficients can be obtained
from the recursion relation, which for $N=1$ is

\begin{equation}\begin{gathered}
\label{3.1}c_{\alpha,-3(l+1)-1}=(3l+2)(3l+1)c_{\alpha,-3l-1}-
2\sum_{m=0}^{l}\sum_{n=0}^{m}
c_{\alpha,-3n-1}\\
c_{\alpha,-3(m-n)-1}c_{\alpha,-3(l-m)-1},\quad l\geq1.
\end{gathered}\end{equation}

In the general case the recursion relation has the similar structure
(see \eqref{2.1} and \eqref{2.8}):

\begin{equation}\begin{gathered}
\label{3.2}c_{\alpha,-(2N+1)(l+1)-1}=((2N+1)l+2N)\ldots((2N+1)l+1)
c_{\alpha,-(2N+1)l-1}\\
+\ldots +\frac{(-1)^N2^{N}(2N-1)!!}{N!}
\sum_{k_1=0}^{l}\sum_{k_2=0}^{k_1}\ldots\sum_{k_{2N}=0}^{k_{2N-1}}
c_{\alpha,-(2N+1)k_{2N}-1}\\
c_{\alpha,-(2N+1)(k_{2N-1}-k_{2N})-1}\ldots
c_{\alpha,-(2N+1)(l-k_1)-1},\quad l\geq1.
\end{gathered}\end{equation}

Suppose that $c_{\alpha,-(2N+1)m-1}$ is a polynomial in $\alpha$ of
degree $(2Nm+1)$ $(0<m\leq l)$. Then from \eqref{3.2} we see that
$c_{\alpha,-(2N+1)(l+1)-1}$ is also a polynomial in $\alpha$ and
$\deg(
c_{\alpha,-(2N+1)(l+1)-1})=2k_{2N}N+1+2(k_{2N-1}-k_{2N})N+1+\ldots
+2(l-k_1)N+1=2Nl+2N+1=2N(l+1)+1$. Q.E.D.
\end{proof}

\begin{theorem}
\label{T:3.1} The following statements are true:

1. at given $n\geq2$ the functions $s_{n,m}^{(N)}$
$(m>p_{n,N}=n(n+1)/2)$ do not contain any new information about the
roots of $Q_n^{(N)}(z)$;

2. $s_{n,(2N+1)\,l}^{(N)}$ is a polynomial in $n$ of degree
$2(Nl+1)$.
\end{theorem}
\begin{proof}
The first statement of the theorem immediately follows from the
correlation

\begin{equation}
\label{3.3}s_{n,m}+s_{n,m-1}A_{n,1}+\ldots
+s_{n,m-p_n}A_{n,p_n}=0,\quad m>p_{n},\quad m\in \textbf{N}
\end{equation}
and the expression \eqref{2.23} (the index $N$ is omitted). Now let
us prove the second statement. From \eqref{2.13} and Lemma
\eqref{L:3.1} we see that in order to find $s_{n,(2N+1)l}^{(N)}$ one
should calculate finite amount of sums $\sum\limits_{i=1}^{n}i^m$,
$m\in \textbf{N}$, $\max m=2Nl+1$. Such sum is computable. And the
result is a polynomial in $n$ of degree $m+1$. This completes the
proof.
\end{proof}

Using the remark at the end of the previous section we obtain that
the roots of $Q_n^{(N)}(z)$ lie on circles with center the origin.
The radii of the circles are equal to $(2N+1)$th roots of the
absolute values of the non-zero roots of $q_n^{(N)}(\xi)$.
Furthermore there are $2N+1$ equally spaced roots of $Q_n^{(N)}(z)$
on a circle and for the real roots of $q_n^{(N)}(\xi)$ and $2(2N+1)$
roots of $Q_n^{(N)}(z)$ ($2N+1$ complex conjugate pairs) are located
on a circle for the complex roots of $q_n^{(N)}(\xi)$.

\section{Coefficients of the Yablonskii - Vorob'ev polynomials}

In the next sections we will find first several coefficients of the
Yablonskii - Vorob'ev polynomials for some members of the $P_2$
hierarchy. While now let us study the general case and prove a
theorem.
\begin{theorem}
\label{T:4.1} The coefficient $A_{n,(2N+1)l}^{(N)}$ is a polynomial
in $n$ of degree $2(N+1)l$. Moreover
$A_{n,(2N+1)l}^{(N)}/T_{n,l}^{(N)}$ is also a polynomial in $n$,
where $T_{n,l}^{(N)}=n(n-1)(n-2)\ldots (n-(n_0-1))$ and
$n_0=\,]\sqrt{1/4+2(2N+1)l}-1/2[$ with $]x[$ denoting $x$ if $x$ is
an integer and $[x]+1$ otherwise.
\end{theorem}
\begin{proof}
The first part of the theorem can be proved by induction on $l$.
Indeed in the case $l=1$ we see that
$A_{n,(2N+1)}=-s_{n,(2N+1)}/(2N+1)$. (Here and up to the end of the
proof the upper index $N$ is omitted.) Consequently using the second
statement of theorem \eqref{T:3.1} we get the correlation $\deg
A_{n,(2N+1)}=\deg s_{n,(2N+1)}=2(N+1)$. Let $A_{n,(2N+1)m}$ be a
polynomial in $n$ of degree $2(N+1)m$ $(m < l)$. The coefficient
$A_{n,(2N+1)l}$ can be found with a help of \eqref{2.25}. Analyzing
this expression we understand that the term
$s_{n,(2N+1)}A_{n,(2N+1)\,l-(2N+1)}$ gives the greatest contribution
into the degree of $A_{n,(2N+1)l}$. Hence $\deg
A_{n,(2N+1)l}=2(N+1)+2(N+1)(l-1)=2(N+1)l$.

In order to prove the second part of the theorem first of all we
should note that $\sum\limits_{i=1}^{n}i^m/n$ is a polynomial in $n$
for all $m\in \textbf{N}\cup\{0\}$. Next let us regard the left-hand
side of \eqref{3.3} also as a polynomial in $n$

\begin{equation}
\begin{gathered}
\label{4.1}R_{n,(2N+1)l}\stackrel{def}{=}s_{n,(2N+1)l}+s_{n,(2N+1)(l-1)}A_{n,2N+1}+\ldots
+s_{n,(2N+1)(l-k)}\\
\times A_{n,(2N+1)k}, \quad k=\max\limits_{(2N+1)j\leq p_n}j,\quad
(2N+1)l>p_n,\quad l\in \textbf{N}.
\end{gathered}
\end{equation}

As far as \eqref{3.3} holds, then $R_{n_0-1,(2N+1)l}=0$, where $n_0$
is obtained from $n_0=\min\limits_{n(n+1)/2\geq (2N+1)l}n$. At the
same time $A_{n,(2N+1)\tilde{l}}$ is equal to
$R_{n,(2N+1)\tilde{l}}$ accurate to a numerical parameter
$(\tilde{l}=\min\limits_{(2N+1)j> n_0(n_0-1)/2}j)$. Continuing in
such a way we complete the proof.
\end{proof}

\section{The Yablonskii - Vorob'ev polynomials associated with rational solutions of $P_2$}

In this section our interest is in the polynomials $Q_n^{(1)}(z)$
associated with the equation \eqref{1.1}. Later in this section the
upper index will be omitted. The power expansion at infinity
corresponding to the solutions of $P_2\equiv P_2^{(1)}$ is the
following \cite{Bruno03}

\begin{equation}
\label{5.1}w(z;\alpha)=-\frac{\alpha}{z}+\sum_{l=1}^{\infty}c_{\alpha,-3l-1}z^{-3l-1},
\quad z\rightarrow \infty,
\end{equation}
where $c_{\alpha,-3l-1}$ $(l>1)$ satisfy the recurrence relation
\eqref{3.1}. Using theorems \eqref{T:3.1} and \eqref{T:4.1} we
obtain

\begin{equation}
\label{5.2}s_{n,m}=\left\{
\begin{gathered}
 0,\quad m\in \textbf{N}\,/\,\{3l,\quad
l\in
\textbf{N}\};\\
S_{2(l+1)}(n),\quad m=3l,\,l\in \textbf{N}.
\end{gathered}
\right.
\end{equation}

\begin{equation}
\label{5.3}A_{n,m}=\left\{
\begin{gathered}
\hfill 0,\quad m\in\{1,2,\ldots
,n(n+1)/2\}\,/\,\{3l,\quad l\in \textbf{N}\};\\
\mathcal{A}_{4l}(n),\quad m=3l,\quad l\in \textbf{N},\quad 3l\leq
n(n+1)/2.
\end{gathered}
\right.
\end{equation}

Here $S_{2(l+1)}(n)$, $\mathcal{A}_{4l}(n)$ are polynomials in $n$
of degree $2(l+1)$ and $4l$, accordingly. Each polynomial $Q_n(z)$
is a monic polynomial of degree $n(n+1)/2$ and can be written as

\begin{equation}
\label{5.4}Q_n(z)=\sum_{l=0}^{[n(n+1)/6]}A_{n,3l}z^{n(n+1)/2\,\,-\,\,3l},\qquad
A_{n,0}=1.
\end{equation}

Hence the polynomial $Q_n(z)$ is divisible by $z$ if and only if $n
\,mod\,3=1$. Besides that $Q_n(z)$ is a polynomial in $z^3$ if
$n\,mod\,3 \neq1$ and $Q_n(z)/z$ is a polynomial in $z^3$ if
$n\,mod\,3 =1$. Since every polynomial $Q_n(z)$ has simple roots and
two successive polynomials do not have a common root then the
rational solution $w(z;n)$ of $P_2$ has $n(n-1)/2$ poles with
residue $1$, which are the roots of $Q_{n-1}(z)$ and $n(n+1)/2$
poles with residue $-1$, which in their turn are the roots of
$Q_n(z)$.

With a help of the expression \eqref{2.5} we can obtain the
symmetric functions of the roots $s_{n,3l}$. The first few of them
are the following

\begin{equation}\begin{gathered}
\label{k1}s_{{{n,3}}}=-\frac12\,n \left( {n}^{2}-1 \right) \left(
n+2 \right),
\end{gathered}\end{equation}

\begin{equation}\begin{gathered}
\label{k2}s_{{{n,6}}}=2\,n \left( {n}^{2}-1 \right)  \left( n+2
\right)
 \left( {n}^{2}+n-5 \right),
\end{gathered}\end{equation}

\begin{equation}\begin{gathered}
\label{k3}s_{{{n,9}}}=-4\,n \left( {n}^{2}-1 \right)  \left( n+2
\right)
 \left( {n}^{2}+n-7 \right)  \left( 3\,{n}^{2}+3\,n-20 \right),
\end{gathered}\end{equation}

\begin{equation}\begin{gathered}
\label{k4}s_{{{n,12}}}=8\,n \left( {n}^{2}-1 \right)  \left( n+2
\right)
 (11\,{n}^{6}+33\,{n}^{5}-259\,{n}^{4}-573\,{n}^{3}\\
 +2348\,{n}^{2
}+2640\,n-7700),
\end{gathered}\end{equation}

\begin{equation}\begin{gathered}
\label{k5}s_{{{n,15}}}=-8\,n \left( {n}^{2}-1 \right)  \left( n+2
\right)
 (91\,{n}^{8}+364\,{n}^{7}-3468\,{n}^{6}-11678\,{n}^{5}+\\
 57138\,{ n}^{4}+134164\,{n}^{3}-454161\,{n}^{2}-523250\,n+1401400);
\end{gathered}\end{equation}

The coefficients $A_{n,3l}$ of the polynomials $Q_n(z)$ can be found
using the expression \eqref{2.24} and the symmetric functions
$s_{n,3l}$. The first few of them are written out below

\begin{equation}\begin{gathered}
\label{k11}A_{{{n,3}}}=\frac{n}{6}\,\left( {n}^{2}-1 \right) \left(
n+2 \right),
\end{gathered}\end{equation}
\begin{equation}\begin{gathered}
\label{k12}A_{{{n,6}}}={\frac {n}{72}}\,\left( {n}^{2}-1 \right)
\left( {n} ^{2}-4 \right)  \left( n-4 \right)  \left( n+5 \right)
\left( n+3
 \right),
\end{gathered}\end{equation}

\begin{equation}\begin{gathered}
\label{k13}A_{{{n,9}}}={\frac {n}{1296}}\,\left( {n}^{2}-1 \right)
\left( { n}^{2}-4 \right)  \left( {n}^{2}-9 \right)  \left( n+4
\right)\\
 \left( {n}^{4}+2\,{n}^{3}-57\,{n}^{2}-58\,n+1120 \right),
\end{gathered}\end{equation}

\begin{equation}\begin{gathered}
\label{k14}A_{{{n,12}}}={\frac {n}{31104}}\,\left( {n}^{2}-1 \right)
 \left( {n}^{2}-4 \right)  \left( {n}^{2}-9 \right)  \left( {n}^{2}-16
 \right)  \left( n+5 \right)  \\
 \left( {n}^{6}+3\,{n}^{5}-109\,{n}^{4}-
223\,{n}^{3}+5148\,{n}^{2}+5260\,n-110880 \right).
\end{gathered}\end{equation}

\begin{equation}\begin{gathered}
\label{k15}A_{{{n,15}}}={\frac {n}{933120}}\,\left( {n}^{2}-1
\right)
 \left( {n}^{2}-4 \right)  \left( {n}^{2}-9 \right)  \left( {n}^{2}-16
 \right)  \left( n+5 \right) ({n}^{10}+\\
 5\,{n}^{9} -200\,{n}^{8}-
830\,{n}^{7}+18917\,{n}^{6}+59677\,{n}^{5}-1072550\,{n}^{4}-\\
2245540\,{ n}^{3}+35648392\,{n}^{2}+36781248\,n-484323840),
\end{gathered}\end{equation}

Besides that the following estimations can be derived

\begin{equation}\begin{gathered}
\label{new1}A_{{{n,3(l+1)}}}\simeq \frac{n^4}{6(l+1)}
A_{n,3l},\quad n\rightarrow \infty,\quad l\geq0;
\end{gathered}\end{equation}
\begin{equation}\begin{gathered}
\label{new2}A_{{{n,3(l+1)}}}\simeq \frac{l}{(l+1)}\frac{A^{2}_{n,3l}}{A_{n,3(l-1)}}
,\quad n\rightarrow \infty,\quad l\geq1.
\end{gathered}\end{equation}

Analysing the expressions \eqref{k11} -- \eqref{k15} we get

\begin{equation}\begin{gathered}
\label{new3}A_{{{n,3l}}}= \frac{n(n^2-1)(n+2)}{6^ll!}
[n^{4l-4)}+2(l-1)n^{4l-5}
-5(l-1)(2l+1)\\
\times n^{4l-6} +\ldots+2^{l-1}(3l-1)!], \quad(l>1)
\end{gathered}\end{equation}

Consequently calculating the coefficients $c_{\alpha,-3l-1}$ of the
expansion \eqref{5.1} one can construct each polynomial $Q_n(z)$.
Several polynomials $Q_n(z)$ are given in Table \ref{t:5.3}.

\begin{table}[h]%[h]
    \caption{The Yablonskii - Vorob'ev polynomials for $P_2$} \label{t:5.3}
    \center
       \begin{tabular}[pos]{l}
        \hline\\
        $Q_0(z) = 1$,\\
        $Q_1(z) = z$,\\
        $Q_2(z) = z^3 + 4$,\\
$Q_3(z) = z^6 + 20z^3 - 80$,\\
$Q_4(z) = (z^9 + 60z^6 + 11 200)z$,\\
$Q_5(z) = z^{15} + 140z^{12} + 2800z^9 + 78 400z^6 - 313 600z^3- 6
272 000$,\\
$Q_6(z) = z^{21} + 280z^{18} + 18480z^{15} + 627
200z^{12}- 17 248 000z^9 + 1 448 832000z^6$ \\ $\qquad \qquad + 19
317 760 000z^3 - 38 635 520 000$,\\
$Q_7(z)=(z^{27} + 504z^{24} + 75
600z^{21} + 5 174 400z^{18} + 62 092 800z^{15} + 13 039 488
000z^{12}$\\$\qquad \qquad - 828 731 904 000z^9- 49 723 914 240
000z^6 - 3 093 932 441 600 000)z$ \\ \\
        \hline
    \end{tabular}
\end{table}

Substituting the polynomials $Q_{n-1}(z)$ and $Q_n(z)$ into
\eqref{1.2} we obtain the rational solution $w(z;n)$ of $P_2$.
Several of them are

\begin{equation*}\begin{gathered}
\label{k15}
w(z;1)={\large -\frac1z} \hfill\\
w(z;2)=-\,{\frac {2({z}^{3}-2)}{z \left( {z}^{3}+4 \right) }} \hfill\\
w(z;3)=-\,{\frac {3{z}^{2} \left( {z}^{6}+8\,{z}^{3}+160 \right) }{
\left( {z }^{3}+4 \right)  \left( {z}^{6}+20\,{z}^{3}-80 \right) }}
\hfill \\
w(z;4)=-\,{\frac
{4({z}^{15}+50\,{z}^{12}+1000\,{z}^{9}-22400\,{z}^{6}-112000
\,{z}^{3}-224000)}{z \left( {z}^{9}+60\,{z}^{6}+11200 \right)
\left( { z}^{6}+20\,{z}^{3}-80 \right) }} \hfill
\end{gathered}\end{equation*}

\section{The Yablonskii - Vorob'ev polynomials associated
with rational solutions of $P_2^{(2)}$}

In this section we will deal with the polynomials $Q_n^{(2)}(z)$
associated with the forth-order analogue to $P_2$ \eqref{1.4a}.
Again the upper index will be omitted. The power expansion at
infinity corresponding to the solutions of \eqref{1.4a} is the
following \cite{Demina01, Demina02}

\begin{equation}
\label{6.1}w(z;\alpha)=-\frac{\alpha}{z}+\sum_{l=1}^{\infty}c_{\alpha,-5l-1}z^{-5l-1},
\quad z\rightarrow \infty,
\end{equation}
where $c_{\alpha,-5l-1}$ $(l>1)$ can be sequently found. Using the
results of the previous sections we get

\begin{equation}
\label{6.2}s_{n,m}=\left\{
\begin{gathered}
 0,\quad m\in \textbf{N}\,/\,\{5l,\quad
l\in
\textbf{N}\};\\
S_{2(2l+1)}(n),\quad m=5l,\,l\in \textbf{N}.
\end{gathered}
\right.
\end{equation}

\begin{equation}
\label{6.3}A_{n,m}=\left\{
\begin{gathered}
\hfill 0,\quad m\in\{1,2,\ldots
,n(n+1)/2\}\,/\,\{5l,\quad l\in \textbf{N}\};\\
\mathcal{A}_{6l}(n),\quad m=5l,\quad l\in \textbf{N},\quad 5l\leq
n(n+1)/2.
\end{gathered}
\right.
\end{equation}
Here $S_{2(2l+1)}(n)$, $\mathcal{A}_{6l}(n)$ are polynomials in $n$
of degree $2(2l+1)$ and $6l$, accordingly. Being of degree
$n(n+1)/2$ the polynomial $Q_n(z)$ can be written as

\begin{equation}
\label{6.4}Q_n(z)=\sum_{l=0}^{[n(n+1)/10]}A_{n,5l}z^{n(n+1)/2\,\,-\,\,5l},\qquad
A_{n,0}=1.
\end{equation}

Therefore the polynomial $Q_n(z)$ is a polynomial in $z^5$ if
$n\,mod\,5 =0$ or $n\,mod\,5 =4$, $Q_n(z)/z$ is a polynomial in
$z^5$ if $n\,mod\,5 =1$ or $n\,mod\,5 =3$, and $Q_n(z)/z^3$ is a
polynomial in $z^5$ if $n\,mod\,5 =2$.

The symmetric functions of the roots $s_{n,5l}$  can be obtained
with a help of the expression \eqref{2.5}. The first few of them are
the following

\begin{equation}\begin{gathered}
\label{kk1}s_{{{n,5}}}=n \left( {n}^{2}-1 \right) \left( {n}^{2}-4
\right)
 \left( n+3 \right),
\end{gathered}\end{equation}

\begin{equation}\begin{gathered}
\label{kk2}s_{{{n,10}}}=6\,n \left({n}^{2}-1 \right)\left( {n}^{2}-4
\right)\left( n+3
\right)(3\,{n}^{4}+6\,{n}^{3}-73\,{n}^{2}-76\,n+504),
\end{gathered}\end{equation}

\begin{equation}\begin{gathered}
\label{kk3}s_{{{n,15}}}=36\,n \,\left( {n}^{2}-1 \right) \left(
{n}^{2}-4 \right)  \left( n+3 \right)\,\, (
15\,{n}^{8}+60\,{n}^{7}-1010\,{n} ^{6} \\
-3240\,{n}^{5} +28759\,{n}^{4}+62988\,{n}^{3}-388124\,{n}^{2}-
420168\,n+2018016 ).
\end{gathered}\end{equation}

Using the expression \eqref{2.24} and the symmetric functions
$s_{n,5l}$ we find the coefficients $A_{n,5l}$ of the polynomials
$Q_n(z)$. The first few of them are

\begin{equation}\begin{gathered}
\label{kk11}A_{{{n,5}}}=-\frac{n}{5}\left( {n}^{2}-1 \right) \left(
{n}^{2}-4
 \right)  \left( n+3 \right),
\end{gathered}\end{equation}

\begin{equation}\begin{gathered}
\label{kk12}A_{{{n,10}}}={\frac {n}{50}}\,\left( {n}^{2}-1 \right)
\left( {n }^{2}-4 \right)  \left( {n}^{2}-9 \right)  \left( n+4
\right)\\
 \left( {n}^{4}+2\,{n}^{3}-85\,{n}^{2}-86\,n+1260 \right),
\end{gathered}\end{equation}

\begin{equation}\begin{gathered}
\label{kk13}A_{{{n,15}}}=-{\frac {n}{750}}\,\left( {n}^{2}-1 \right)
\left( {n}^{2}-4 \right)  \left( {n}^{2}-9 \right)  \left(
{n}^{2}-16
\right)  \left( n+5 \right)\\
({n}^{8}+4\,{n}^{7}-248\,{n}^{6}-
758\,{n}^{5}+26959\,{n}^{4}+55186\,{n}^{3}-1107792\,{n}^{2}-\\
-1135512\,n +15135120).
\end{gathered}\end{equation}

Besides that we obtain the following estimations

\begin{equation}\begin{gathered}
\label{new1}A_{{{n,5(l+1)}}}\simeq -\frac{n^6}{5(l+1)}A_{n,5l},\quad n\rightarrow \infty,
\quad l\geq0;
\end{gathered}\end{equation}
\begin{equation}\begin{gathered}
\label{new2}A_{{{n,5(l+1)}}}\simeq \frac{l}{(l+1)}\frac{A^{2}_{n,5l}}{A_{n,5(l-1)}}
,\quad n\rightarrow \infty,\quad l\geq1.
\end{gathered}\end{equation}

Thus calculating the coefficients $c_{\alpha,-5l-1}$ of the
expansion \eqref{6.1} each polynomial $Q_n(z)$ can be constructed.
Several polynomials $Q_n(z)$ are gathered in Table \ref{t:6.3}

\begin{table}[h]%[h]
    \caption{The Yablonskii - Vorob'ev polynomials for $P_2^{(2)}$} \label{t:6.3}
    \center
       \begin{tabular}[pos]{l}
        \hline \\
        $Q_0(z) = 1$,\\
        $Q_1(z) = z$,\\
        $Q_2(z) = z^3$,\\
$Q_3(z) = z(z^5-144)$\\ $Q_4(z) = z^{10}-1008z^5 -48 384$,\\
$Q_5(z) = z^{15}- 4032z^{10} - 3 048 192z^5 + 146 313 216$,\\
$Q_6(z) =z(z^{20}- 12 096z^{15} - 21 337 344z^{10}- 33 798 352
896z^5
 - 4 866 962 817 024)$, \\
$Q_7(z)=z^3(z^{25}- 30 240z^{20}- 55 883 520z^{15}- 1 182 942 351
 360z^{10}
+701 543 488 297 107 456)$\\ \\
        \hline
    \end{tabular}
\end{table}

Substituting the polynomials $Q_{n-1}(z)$ and $Q_n(z)$ into
\eqref{1.2} yields the rational solution $w(z;n)$ of $P_2^{(2)}$.
Several of them are

\begin{equation*}\begin{gathered}
\label{k15}
w(z;1)={\large -\frac1z} \hfill\\
w(z;2)=-\,\frac {2}{z} \hfill\\
w(z;3)=-\,{\frac {3({z}^{5}+96)}{z \left( {z}^{5}-144 \right) }}
\hfill \\
w(z;4)=-\,{\frac {4({z}^{15}-72\,{z}^{10}+217728\,{z}^{5}-1741824)}{
z\left( {z} ^{5}-144 \right) \left( {z}^{10}-1008\,{z}^{5}-48384
\right) }} \hfill \\
w(z;5)=-\,{\frac {5{z}^{4} \left(
{z}^{20}-2016\,{z}^{15}+6967296\,{z}^{10}+
97542144\,{z}^{5}+294967443456 \right) }{ \left(
{z}^{15}-4032\,{z}^{ 10}-3048192\,{z}^{5}+146313216 \right)  \left(
{z}^{10}-1008\,{z}^{5}- 48384 \right) }} \hfill
\end{gathered}\end{equation*}

\section{The Yablonskii - Vorob'ev polynomials associated with
rational solutions of $P_2^{(3)}$}

In this section we will briefly review the case of the polynomials
$Q_n^{(3)}(z)$ associated with the rational solutions of the
sixth-order analogue to $P_2$

\begin{equation}\begin{gathered}
\label{7.1}w_{zzzzzz}-14\,{w}^{2}w_{{{zzzz}}}-56\,ww_{{z}}w_{{{
zzz}}}-42\,w(w_{{{zz}}})^{2}-70\,(w_{{z}})^{2}w_{{{zz}}}\\
+70\,{w}^{4}w_{{{zz}}}
+140\,{w}^{3}(w_{{z}})^{2}-20\,{w}^{7}-z\,w-\alpha=0.
\end{gathered}
\end{equation}

Later the upper index will be omitted. The power expansion at
infinity corresponding to the solutions of \eqref{7.1} can be
written as

\begin{equation}
\label{7.2}w(z;\alpha)=-\frac{\alpha}{z}+\sum_{l=1}^{\infty}c_{\alpha,-7l-1}z^{-7l-1},
\quad z\rightarrow \infty,
\end{equation}
where $c_{\alpha,-7l-1}$ $(l>1)$ are sequently found. Again using
the results of the previous sections we obtain

\begin{equation}
\label{7.3}s_{n,m}=\left\{
\begin{gathered}
 0,\quad m\in \textbf{N}\,/\,\{7l,\quad
l\in
\textbf{N}\};\\
S_{2(3l+1)}(n),\quad m=7l,\,l\in \textbf{N}.
\end{gathered}
\right.
\end{equation}

\begin{equation}
\label{7.4}A_{n,m}=\left\{
\begin{gathered}
\hfill 0,\quad m\in\{1,2,\ldots
,n(n+1)/2\}\,/\,\{7l,\quad l\in \textbf{N}\};\\
\mathcal{A}_{8l}(n),\quad m=7l,\quad l\in \textbf{N},\quad 7l\leq
n(n+1)/2.
\end{gathered}
\right.
\end{equation}
Here $S_{2(3l+1)}(n)$, $\mathcal{A}_{8l}(n)$ are polynomials in $n$
of degree $2(3l+1)$ and $8l$, accordingly. Since the polynomial
$Q_n(z)$ is of degree $n(n+1)/2$, then it can be presented in the
form

\begin{equation}
\label{7.5}Q_n(z)=\sum_{l=0}^{[n(n+1)/14]}A_{n,7l}z^{n(n+1)/2\,\,-\,\,7l},\qquad
A_{n,0}=1.
\end{equation}

We see that the polynomial $Q_n(z)$ is a polynomial in $z^7$ if
$n\,mod\,7 =0$ or $n\,mod\,7 =6$, $Q_n(z)/z$ is a polynomial in
$z^7$ if $n\,mod\,7 =1$ or $n\,mod\,7 =5$, $Q_n(z)/z^3$ is a
polynomial in $z^7$ if $n\,mod\,7 =2$ or $n\,mod\,7 =4$ and
$Q_n(z)/z^6$ is a polynomial in $z^7$ if $n\,mod\,7 =3$.

With a help of the expressions \eqref{2.5} the symmetric functions
of the roots $s_{n,7l}$ can be obtained. The first few of them are
the following

\begin{equation}\begin{gathered}
\label{kkk1}s_{{{n,7}}}=-\frac{5n}{2} \left( {n}^{2}-1 \right)
\left( {n}^{2}-4
 \right)  \left( {n}^{2}-9 \right)  \left( n+4 \right),
\end{gathered}\end{equation}

\begin{equation}\begin{gathered}
\label{kkk2}s_{{{n,14}}}=40\,n \left( {n}^{2}-1 \right) \left(
{n}^{2}-4
 \right)  \left( {n}^{2}-9 \right)  \left( n+4 \right) (5\,{n}^
{6}+15\,{n}^{5}\\
-340\,{n}^{4}-705\,{n}^{3}+8651\,{n}^{2}+9006\,n-77220),
\end{gathered}\end{equation}

\begin{equation}\begin{gathered}
\label{kkk3}s_{{{n,21}}}=-800\,n \left( {n}^{2}-1 \right) \left(
{n}^{2}-4
 \right)  \left( {n}^{2}-9 \right)  \left( n+4 \right)(35\,{n}
^{12}+210\,{n}^{11}\\
-6860\,{n}^{10}-36225\,{n}^{9}+629265\,{n}^{8}+
2736720\,{n}^{7}-32792630\,{n}^{6}\\
-108110865\,{n}^{5}+989372966\,{n}^{
4}+2162197152\,{n}^{3}\\
-16042160664\,{n}^{2}-17141744880\,n+ 107749699200).
\end{gathered}\end{equation}

Using the expression \eqref{2.24} and the symmetric functions
$s_{n,7l}$ we find the coefficients $A_{n,7l}$ of the polynomials
$Q_n(z)$

\begin{equation}\begin{gathered}
\label{kkk11}A_{{{n,7}}}={\frac {5n}{14}} \left( {n}^{2}-1 \right)
\left( {n} ^{2}-4 \right)  \left( {n}^{2}-9 \right)  \left( n+4
\right),
\end{gathered}\end{equation}

\begin{equation}\begin{gathered}
\label{kkk12}A_{{{n,14}}}={\frac {5n}{392}}\left( {n}^{2}-1 \right)
\left( { n}^{2}-4 \right)  \left( {n}^{2}-9 \right)  \left(
{n}^{2}-16 \right)
 \left( n+5 \right)\\
 (5\,{n}^{6}+15\,{n}^{5}-1105\,{n}^{4}-2235
\,{n}^{3}+56540\,{n}^{2}+57660\,n-864864),
\end{gathered}\end{equation}

\begin{equation}\begin{gathered}
\label{kk13}A_{{{n,21}}}={\frac {25n}{16464}} \left( {n}^{2}-1
\right)
 \left( {n}^{2}-4 \right)  \left( {n}^{2}-9 \right)  \left( {n}^{2}-16
 \right)  \left( {n}^{2}-25 \right)
 \left( n+6 \right) \\
(5{n} ^{12}+30{n}^{11}-3235{n}^{10}-16450{n}^{9}+985395{n}^{8}+
4040610{n}^{7}\\
-137483057{n}^{6}-426660726{n}^{5}+9606229564{n}
^{4}+19928307448{n}^{3}\\
-331282163616\,{n}^{2}-341318105856\,n+ 4505374089216).
\end{gathered}\end{equation}

Besides that the following estimations can be derived

\begin{equation}\begin{gathered}
\label{new1}A_{{{n,7(l+1)}}}\simeq \frac{5n^8}{14(l+1)}
A_{n,7l},\quad n\rightarrow \infty,\quad l\geq0;
\end{gathered}\end{equation}
\begin{equation}\begin{gathered}
\label{new2}A_{{{n,7(l+1)}}}\simeq \frac{l}{(l+1)}\frac{A^{2}_{n,7l}}{A_{n,7(l-1)}}
,\quad n\rightarrow \infty,\quad l\geq1.
\end{gathered}\end{equation}

Thus calculating the coefficients $c_{\alpha,-7l-1}$ of the
expansion \eqref{7.2} each polynomial $Q_n(z)$ can be constructed.
Several polynomials $Q_n(z)$ are gathered in Table \ref{t:7.3}

\begin{table}[h]%[h]
    \caption{The Yablonskii - Vorob'ev polynomials for $P_2^{(3)}$} \label{t:7.3}
    \center
       \begin{tabular}[pos]{l}
        \hline \\
        $Q_0(z) = 1$,\\
        $Q_1(z) = z$,\\
        $Q_2(z) = z^3$,\\
$Q_3(z) = z^6$,\\
$Q_4(z) = z^3(z^7 + 14 400)$,\\
$Q_5(z) = z(z^{14} + 129 600z^7 - 373 248 000)$,\\
$Q_6(z) =z^{21} + 648 000z^{14}- 24 634 368 000z^7 -35 473 489 920
000$,
\\$Q_7(z)=z^{28} + 2 376 000z^{21}- 825 251 328 000z^{14}- 30 436 254 351 360 000z^7
$\\$\qquad \qquad +43 828 206 265 958 400 000$ \\ \\
        \hline
    \end{tabular}
\end{table}

Substituting the polynomials $Q_{n-1}(z)$ and $Q_n(z)$ into
\eqref{1.2} yields the rational solution $w(z;n)$ of $P_2^{(3)}$.
Several of them are

\begin{equation*}\begin{gathered}
\label{k15}
w(z;1)={\large -\frac1z} \hfill\\
w(z;2)=-\,\frac {2}{z} \hfill\\
w(z;3)=-\,{\frac {3}{z}}
\hfill \\
w(z;4)=-\,{\frac {4({z}^{7}-5400)}{z \left( {z}^{7}+7200 \right) }} \hfill \\
w(z;5)=-\,{\frac
{5({z}^{21}-8640\,{z}^{14}+634521600\,{z}^{7}+268738560000)}{z
 \left( {z}^{14}+64800\,{z}^{7}-93312000 \right)  \left( {z}^{7}+7200
 \right) }} \hfill
\end{gathered}\end{equation*}

\section {Conclusion}

In this paper a method for constructing the polynomials associated
with rational solutions of the $P_2$ hierarchy has been presented.
The basic idea of the method is to use power expansions for
solutions of the equations studied. These power expansions can be
obtained with a help of algorithms of power geometry \cite{Bruno01,
Bruno02}. Using our approach we have found: the degree of each
polynomial, formulas for its coefficients, correlations between its
roots. Besides that we have established the structure of the
polynomials. The main computations have been done simultaneously for
all members of the hierarchy.  Our method is sufficiently general
and can be applied for constructing polynomials associated with
rational or algebraic solutions of other nonlinear differential
equations \cite{Okamoto01, Clarkson04, Kajiwara01, Nuomi01,
Umemura01}.

\section {Acknowledgments}

This work was supported by the International Science and Technology
Center under Project B 1213.

\end{document}